\documentclass[prl,twocolumn,showpacs,preprintnumbers,aps,nofootinbib]{revtex4-1}
\usepackage{graphicx}
\usepackage{dcolumn}
\usepackage{bm}
\usepackage{amsmath,amssymb,amsfonts}
\usepackage{latexsym}
\usepackage{color}

\begin{document}


\title{\boldmath
Enhanced $B \to \mu \bar\nu$ Decay at Tree Level
 as Probe of Extra Yukawa Couplings
 }

\author{Wei-Shu Hou, Masaya Kohda, Tanmoy Modak and Gwo-Guang Wong}
\affiliation{Department of Physics, National Taiwan University, Taipei 10617, Taiwan}
\bigskip

\date{\today}

\begin{abstract}
With no New Physics seen at the LHC,
a second Higgs doublet remains attractive and plausible.
The ratio ${\cal R}_B^{\mu/\tau}
 = {\cal B}(B \to \mu \bar\nu)/{\cal B}(B \to \tau \bar\nu)$
is predicted at 0.0045 in both the Standard Model and 
the two Higgs doublet model {type II}, but it can differ 
if extra Yukawa couplings exist in Nature. 
Considering recent Belle update on $B \to \mu\bar\nu$, 
the ratio could be up by a factor of two 
in the general two Higgs doublet model, 
which can be 
probed by the Belle~II experiment with just a few ab$^{-1}$.
\end{abstract}


\maketitle

\paragraph{Introduction.---}

Despite observing 
the 125~GeV boson $h$ at
the Large Hadron Collider (LHC), no New Physics (NP)
 beyond the Standard Model (SM) has so far emerged.
%
Supersymmetry (SUSY) is not seen, nor extra gauge bosons
or any new particle. Except for weaker couplings,
any new symmetry 
must be at rather high scale.
In this light, we ought to reexamine the validity
of any {\it presumed} symmetry. 
In this Letter we reflect on the 
$Z_2$ symmetry
imposed on two Higgs doublet models (2HDM) to 
satisfy 
Natural Flavor Conservation  (NFC)~\cite{Glashow:1976nt}.

Although extra Higgs bosons remain elusive, 
the existence of one doublet 
makes 
a second doublet 
plausible.
The most popular 2HDM type~II~{\cite{Branco:2011iw}}, 
where $u$- and $d$-type quarks receive mass
from separate doublets, is automatic in SUSY.
The NFC condition~\cite{Glashow:1976nt} 
removes flavor changing neutral Higgs (FCNH) couplings 
by demanding each mass matrix comes from 
only one 
Yukawa matrix.
But this removes {\it all} extra Yukawa couplings 
that naturally should exist in a 2HDM!
But with discovery of $m_h < m_t$, 
it was stressed~\cite{Chen:2013qta} that 
FCNH, or extra Yukawa couplings in general, is an experimental issue,
as attested by the ATLAS and CMS pursuit~\cite{Aaboud:2018oqm}
 of $t\to ch$, $uh$, and 
augmented by the hint~\cite{Khachatryan:2015kon} 
of $h \to \tau\mu$ in Run~1 data of CMS,
although it disappeared with Run~2 data~\cite{Sirunyan:2017xzt}.
In this Letter, we explore the issue of extra Yukawa couplings and 
propose 
$\Gamma(B \to \mu\bar\nu)/\Gamma(B \to \tau\bar\nu)$, 
or $R_B^{\mu/\tau}$, as a 
leading probe at the upcoming Belle~II experiment. 

In the LHC era, a handful of so-called ``flavor anomalies''~\cite{Hou:2019dgh} 
do exist, but the recent trend 
is ``softening''.
Belle announced a new measurement~\cite{Abdesselam:2019dgh}
of the $R_D$ and $R_{D^*}$ ratios,
or ${\cal B}(B \to D^{(*)}\tau\nu)/{\cal B}(B \to D^{(*)}\mu\nu)$, 
using semileptonic tag, 
finding consistency with SM,
and the world average tension with SM expectation
decreases from 3.8$\sigma$ to 3.1$\sigma$.
For the clean $R_K$ ratio,
${\cal B}(B \to K\mu^+\mu^-)/{\cal B}(B \to Ke^+e^-)$,
using Run~2 data taken in 2016 
{alone, LHCb finds~\cite{Aaij:2019wad}
consistency with SM, 
but is at 1.9$\sigma$~\cite{Humair} with reanalyzed Run~1 result,
although the combined data is still at 2.5$\sigma$ from SM.}
In any case, these ``anomalies'' need confirmation with full Run~2 data,
as well as Belle~II scrutiny. 
%
The $R_B^{\mu/\tau}$ ratio adds to the list,
and could be an early NP probe at Belle~II.

The recent Belle remeasurement~\cite{Belle-B2munu_Morio19},
\begin{equation}
 {\cal B}(B\to\mu\bar\nu) = (5.3 \pm 2.0 \pm 0.9) \times 10^{-7},
 \;\; ({\rm Belle~2019})
\label{eq:B2munu-Belle19}
\end{equation}
supersedes the published 
$(6.46 \pm 2.22 \pm 1.60) \times 10^{-7}$~\cite{Sibidanov:2017vph}.
The central value dropped slightly, but the improved systematics 
moves the significance up from 2.4$\sigma$ to 2.8$\sigma$.
The $B \to \ell \bar\nu_\ell$ decay branching fraction 
in SM is 
\begin{align}
 {\cal B}(B \to \ell \bar\nu_\ell)|^{\rm SM}  = |V_{ub}|^2 f_{B}^2
      \frac{G_F^2 m_\ell^2 m_{B}}{8\pi\Gamma_{B}}
      \Bigl(1 - \frac{m_\ell^2}{m_{B}^2}\Bigr)^2,
\label{eq:BqellnuSM}
\end{align}
where helicity suppression by $m_\ell^2/m_{B}^2$ makes it
quite rare, but also more susceptible to NP effects.
Using $f_{B} = 190$ MeV from FLAG~\cite{Aoki:2019cca}, 
and exclusive value $|V_{ub}|^{\rm excl.} = 3.70 \times 10^{-3}$
 from PDG~\cite{Tanabashi:2018oca}, we find 
${\cal B}(B \to \ell \bar\nu_\ell)|^{\rm SM} \simeq 3.92 \times 10^{-7}$.
Eq.~(\ref{eq:B2munu-Belle19}) allows a mild enhancement, echoing
an old result from BaBar~\cite{Aubert:2009ar}.

%
The effect in 2HDM~II is also well known~\cite{Hou:1992sy},
\begin{align}
 {\cal B}(B \to \ell \bar\nu_\ell)|^{\rm 2HDM\,II}
  = r_H \, {\cal B}(B \to \ell \bar\nu_\ell)|^{\rm SM},
\label{eq:Bqellnu2HDM2}
\end{align}
where, with {$m_{H^+}$} the $H^+$ mass and $\tan\beta$ 
the ratio of vacuum expectation values (v.e.v.) of the two doublets,
{\begin{align}
 r_H 
 \cong\biggl(1 -\tan ^2\beta\,\frac{m_{B}^2}{m_{H^+}^2}\biggr)^2.
\label{eq:rH}
\end{align}
}
As $r_H$ is $m_\ell$-independent, as in SM, 2HDM~II gives
\begin{align}
 {\cal R}_{B}^{\mu/\tau} \equiv
   \frac{{\cal B}(B \to \mu \bar\nu)}{{\cal B}(B \to \tau \bar\nu)}
   = \frac{m_\mu^2(m_{B}^2 -m_\mu^2)^2}{m_\tau^2(m_{B}^2 -m_\tau^2)^2}
 \cong 0.0045,
\label{eq:RBmutauSM}
\end{align}
as stressed in a recent review~\cite{Chang:2017wpl}.
%
It is also independent of $|V_{ub}|$.
Together with $B \to \tau\bar\nu$ being consistent~\cite{Tanabashi:2018oca} 
with SM, one usually expects
${\cal B}(B \to \mu\bar\nu)$ to be SM-like,
as reflected in the Belle~II Physics Book~\cite{Kou:2018nap}.
%

Belle~II can check 
whether 
${\cal R}_{B}^{\mu/\tau} \cong 0.0045$ holds. 
A deviation would not only be beyond SM, 
but rule out 2HDM~II convincingly.
We will 
show that 
if there exist extra Yukawa couplings, 
${\cal B}(B \to \mu\bar\nu)$ can become enhanced or suppressed, 
while ${\cal B}(B \to \tau\nu)$ would be 
SM-like.
Thus, a 
measurement of ${\cal B}(B\to\mu\bar\nu)$
plus refining
${\cal B}(B\to\tau\bar\nu)$
would open up a probe of extra Yukawa couplings
that complements 
the LHC search for
$t \to ch(uh)$ and $h \to \tau\mu$,
but involving the $H^+$ boson.

\paragraph{Formalism.---}

Compared with Eqs.~(\ref{eq:BqellnuSM}) and (\ref{eq:Bqellnu2HDM2}), 
the $\ell$ index to $\bar\nu$ in Eq.~(\ref{eq:RBmutauSM}) was dropped, 
as the $\bar\nu_\ell$ flavor is not measured, 
but is relevant for 2HDM without $Z_2$.
Called 2HDM~III earlier~\cite{Hou:1991un},
2HDM without $Z_2$ 
followed the Cheng-Sher ansatz~\cite{Cheng:1987rs},
that a trickle-down 
$\sqrt{m_im_j}$ mass-mixing pattern   
may loosen the need for NFC~\cite{Glashow:1976nt} to forbid FCNH. 
The $h$ boson discovery made a discrete symmetry
appear {\it ad hoc}~\cite{Chen:2013qta}:
existence of $tch$ or $h\tau \mu$ FCNH couplings 
should be an experimental question.

For each type of charged fermion $F = u,\, d,\, \ell$,
a second set of Yukawa matrices $\rho_{ij}^F$
comes from a second scalar doublet, where some
trickle-down flavor pattern helps hide the effects,
 in particular from FNCH couplings.
It was revealed recently~\cite{Hou:2017hiw} that
``NFC protection against FCNH can be replaced by 
approximate alignment, together with a flavor organizing 
principle 
reflected in SM itself'', and that
the Cheng-Sher $\sqrt{m_im_j}$ pattern may 
be too strong an assumption.

Approximate alignment emerged 
with LHC~\cite{cos_b-a} Run~1 data:
the $h$ boson appears 
rather close~\cite{Khachatryan:2016vau} to the SM Higgs,
and the two $CP$-even scalars, $h^0$ and $H^0$, 
do not mix much; in 2HDM~II notation, 
the mixing angle $\cos(\alpha -\beta)$ is 
small, which is 
now affirmed by Run~2 data~\cite{Sirunyan:2018koj, {ATLAS:2019slw}}.
But in 2HDM without $Z_2$ (which we now call g2HDM),
$\tan\beta$ is unphysical~{\cite{untbeta}}, hence we 
{replace $\cos(\alpha -\beta)$ by $\cos\gamma$}~\cite{Hou:2017hiw}.
The $tch$ coupling is then $\rho_{tc}\cos\gamma$
 ($\rho_{ct}$ is already constrained 
 to be small~\cite{Chen:2013qta, Altunkaynak:2015twa}),
and $h\tau\mu$ coupling is $\rho_{\tau\mu}\cos\gamma$.
Approximate alignment can
suppress $t\to ch$ or $h \to \tau\mu$ rates,
without invoking tiny $\rho_{tc}$ or $\rho_{\tau\mu}$.

%
%
The fundamental $H^+$ Yukawa couplings 
\begin{align}
 - \bar{u} \bigl( V \rho^d R - \rho^{u\dagger} V L \bigr) d \, H^+
  -\bar{\nu}\bigl( \rho^\ell R \bigr) \ell \, H^+ + {\rm H.c.},
\label{eq:H+Yuk}
\end{align}
are independent of $\cos\gamma$,
where $V$ is the CKM matrix,
${L,\,R} \equiv ( 1\mp \gamma_5 )/2$, and
$u$, $d$, $\ell$ are in matrix notation. 
%
%
They give rise to the branching fraction
\begin{eqnarray}
 && {\cal B}(B \to \ell \bar\nu) = {\cal B}(B \to \ell \bar\nu_\ell)|^{\rm SM} \nonumber\\
 && \quad \times \sum_{\ell' = e,\mu,\tau} \left| \delta_{\ell'\ell}
      - \frac{m_{B}^2 v^2 \rho_{\ell'\ell}^* (\rho_{ib} V_{ui} + \rho_{iu}^* V_{ib})}
                {2V_{ub} m_{{H^+}}^2 m_\ell (m_b + m_u)}\right|^2,
\label{eq:Bqellnu_gen}
\end{eqnarray}
with explicit sum over $\bar\nu_{\ell'}$ flavor,
and sum over $i$ implied.
%
Expanding $\sum_i \rho_{ib} V_{ui} = \rho_{bb} V_{ub}
 +  \rho_{sb} V_{us} +  \rho_{db} V_{ud} \cong \rho_{bb} V_{ub}$,
since 
$\rho_{sb}$ and  $\rho_{db}$ 
are constrained severely at tree level by $B_s$ and $B_d$ meson mixings.
Expanding $\sum_i \rho_{iu}^* V_{ib} = \rho_{tu}^* V_{tb}
 +  \rho_{cu}^* V_{cb} +  \rho_{uu}^* V_{ub} \cong \rho_{tu}^* V_{tb}$,
as 
$\rho_{cu}$ 
is constrained by $D^0$ mixing,
and  $\rho_{uu}$ 
is suppressed by mass-mixing hierarchy,
with both terms 
CKM suppressed.
Note that our result is not affected by the PMNS matrix
in the neutrino sector, so long that it is unitary.

After some rearrangement, the factor becomes
\begin{equation}
 \sum_{\ell' = e,\mu,\tau} \left|  \delta_{\ell'\ell}
        - \frac{m_B^2}{m_{{H^+}}^2}  \frac{\rho_{\ell'\ell}^*}{\lambda_\ell}
          \Bigl( \frac{\rho_{bb}}{\hat m_b}
                 + \frac{\rho_{tu}^*}{\hat m_b} \frac{V_{tb}}{V_{ub}}\Bigr) \right|^2,
\label{eq:Mod}
\end{equation}
where $\lambda_\ell = \sqrt2 m_\ell/v$ is the lepton Yukawa coupling
with $v \cong 246$ GeV,
and $\hat m_b = \sqrt2 m_b/v$ is defined similarly.
But since $m_b$ arises from hadronic matrix element, 
it should be run to $m_{{H^+}}$ scale.
Following PDG~\cite{Tanabashi:2018oca},
we first calculate $\overline{\text{MS}}$ running mass $\overline{m_b}(m_b)$
at pole mass, then evolve to $\mu = m_{{H^+}}$ by
$\overline{m_b}(\mu) = c(\alpha_s(\mu))/c(\alpha_s(m_b)) \, \overline{m_b}(m_b)$,
where 
$c(x)$ is taken with four-loop accuracy 
using 
$\overline{\mbox{MS}}$ three-loop $\alpha_s$ at scale $\mu$ in five-flavor scheme.

{The first notable thing in Eq.~(\ref{eq:Mod}) is
   the $V_{tb}/V_{ub}$ enhancement of the $\rho_{tu}$ effect,
   which does not arise in 2HDM II.}
One can ignore $\rho_{bb}/\hat m_b = {\cal O}(1)$ in g2HDM,
so long that ${\rho_{tu}^*}/{\hat m_b}$ 
does not approach $|V_{ub}/V_{tb}| \sim 0.004$.
Note {also} that, taking
 $\rho_{bb} = -\lambda_b\tan\beta$ and
 $\rho_{\ell'\ell} = - \lambda_\ell\tan\beta \, \delta_{\ell'\ell}$, 
and setting\ $\rho_{tu} = 0$,
one recovers the $r_H$ factor of 2HDM~II.

\begin{figure*}[t]
\center
 \includegraphics[width=6.7cm]{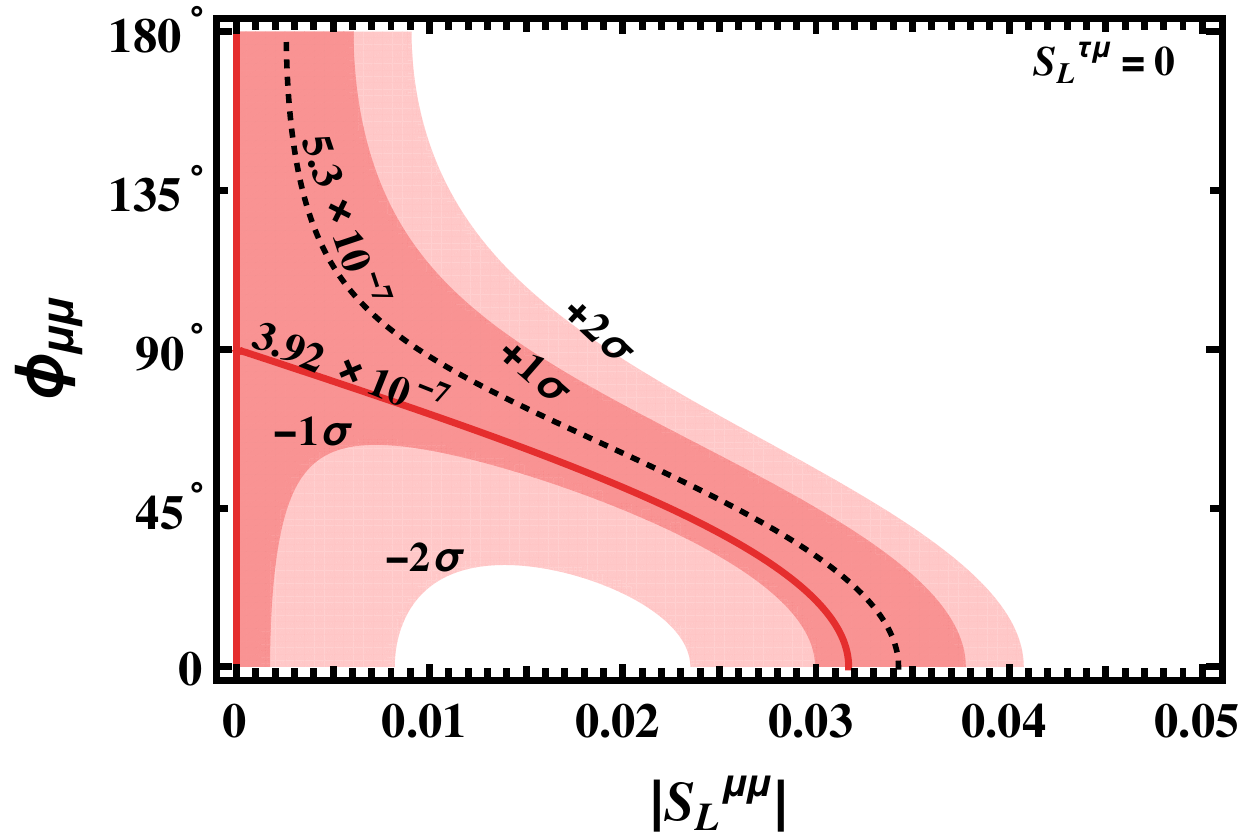} \hskip0.3cm
 \includegraphics[width=6.54cm]{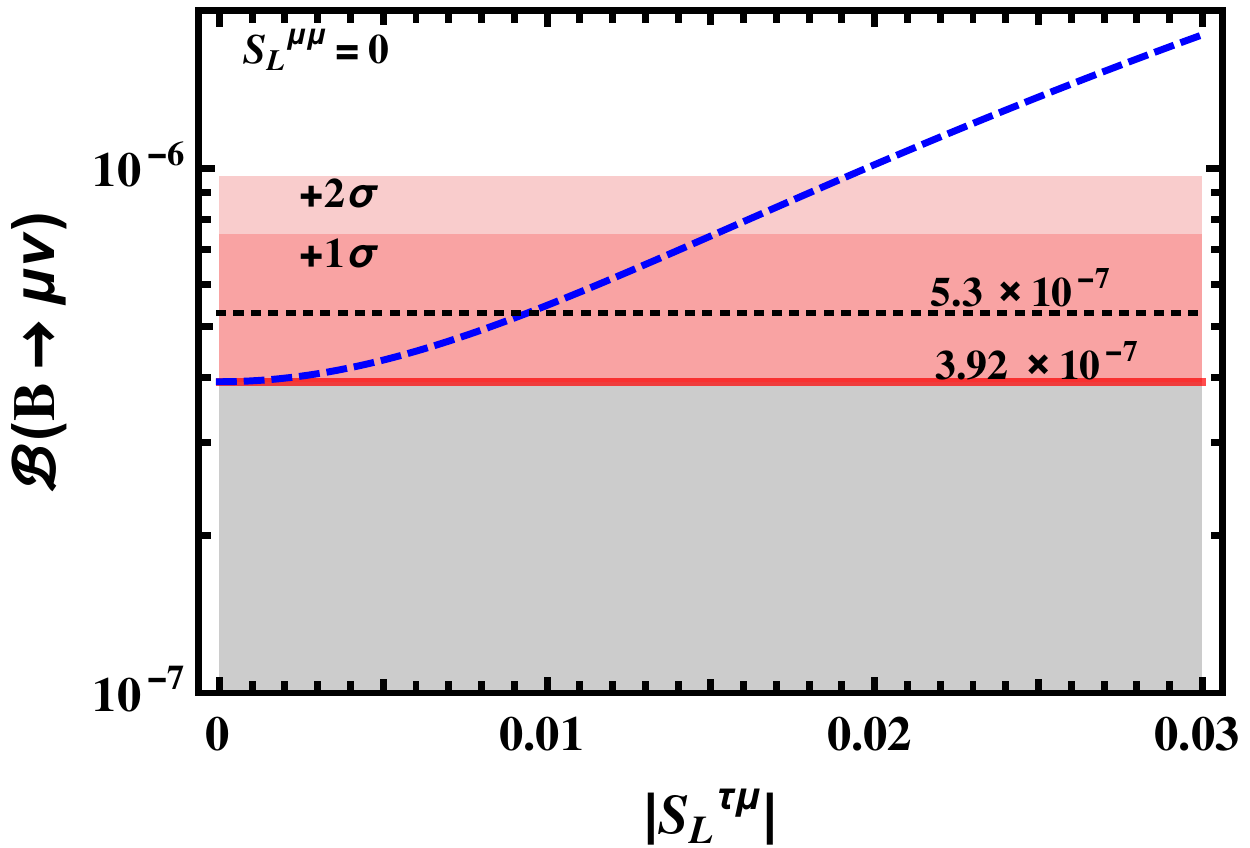}
\caption{
Assuming the other effect in Eq.~(\ref{eq:Bqmunu}) is turned off:
[left]~projection of Belle $B\to \mu\bar\nu$ result~\cite{Belle-B2munu_Morio19}
 in $\bigl | S_L^{\mu\mu} \bigr|$--$\phi_{\mu\mu}$ plane; and
[right]~${\cal B}(B \to \mu\bar\nu)$ vs $| S_L^{\tau\mu}|$ (blue dashed line).
The $\pm 1\sigma$ and $\pm 2\sigma$ allowed regions 
are in dark and light pink shades, 
and black dotted (red solid) line denotes Belle central (SM) value.
The gray shaded region below SM value in right panel 
is theoretically inaccessible. 
The $\overline {\rm MS}$ mass $\overline{m_b}(m_b)$ is used.
}
\label{fig:fig1}
\end{figure*}

\begin{figure*}[t]
\center
\includegraphics[width=6.7cm]{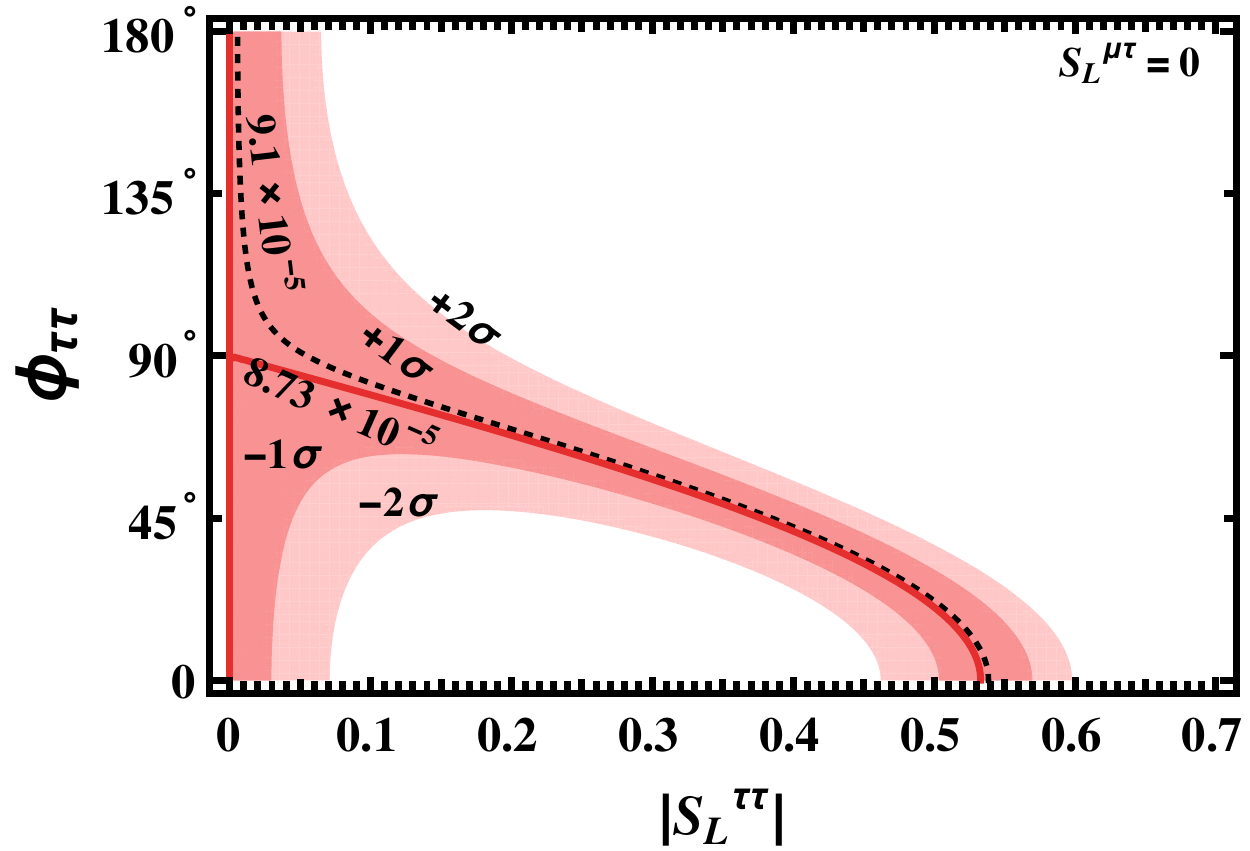} \hskip0.3cm
\includegraphics[width=6.54cm]{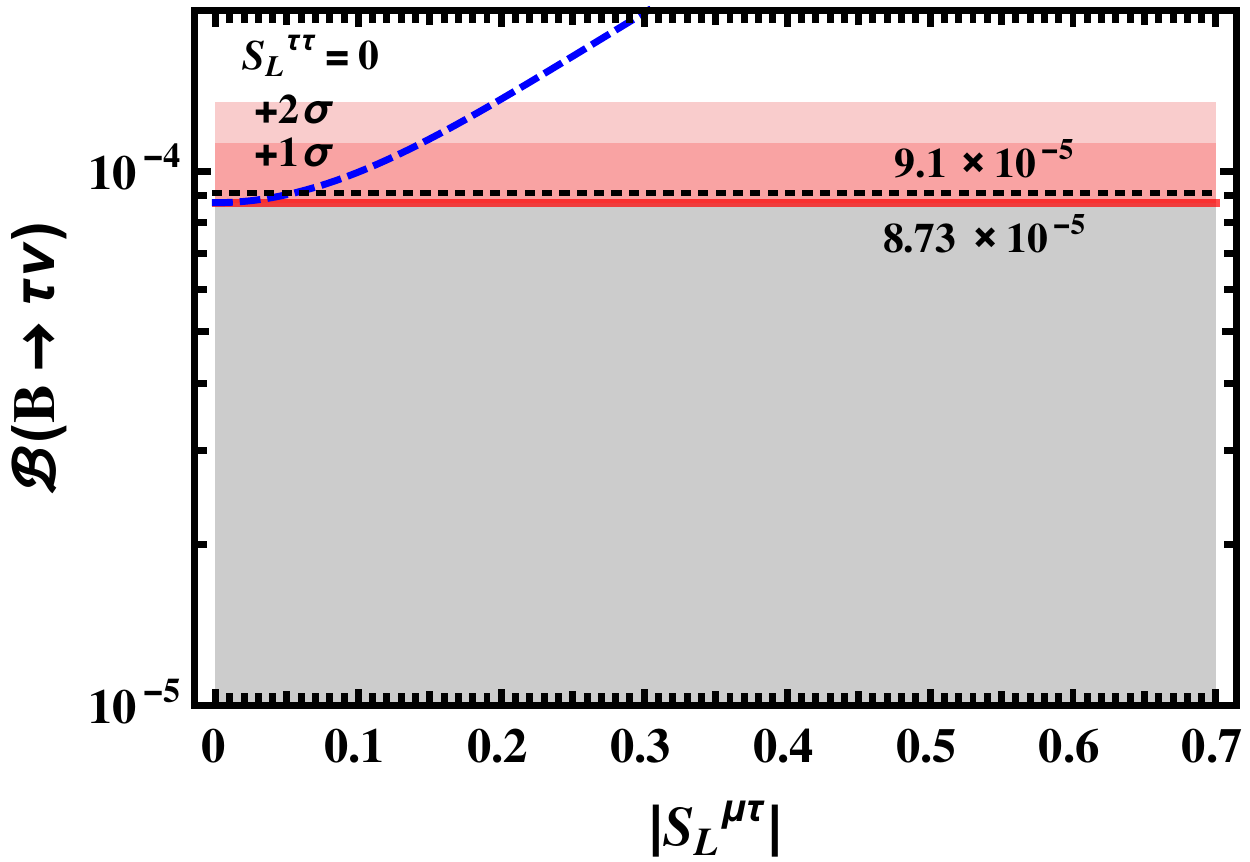} 
\caption{
Similar to Fig. 1, but for $B \to \tau\nu$ decay.
}
\label{fig:fig3}
\end{figure*}

\paragraph{Results.---}

To make contact with experiment, 
we recast in the notation of BaBar's $R_{D^{(*)}}$ paper~\cite{Lees:2013uzd}, 
the form of which the new Belle analysis~\cite{Belle-B2munu_Morio19} has followed, 
\begin{eqnarray}
 {\cal B}(B \to \ell \bar\nu) && = {\cal B}(B \to \ell \bar\nu_\ell)|^{\rm SM} \nonumber\\
 && \times \sum_{\ell' = e,\mu,\tau} \left| \delta_{\ell'\ell}
      - \frac{m_{B}^2}{m_b m_\ell} S_L^{\ell'\ell}\right|^2,
\label{eq:Bqellnu_SL}
\end{eqnarray}
where $S_L^{\ell'\ell}$, at $m_b$ scale, are 
the ratios of NP Wilson coefficients (of 4-Fermi operators) with SM ones.
We note that a $+S_R^{\ell'\ell}$ contribution, proportional to 
$\rho_{bb}$ as seen in Eq.~(\ref{eq:Mod}), is negligible in g2HDM 
because of  the $|V_{tb}/V_{ub}|$ enhancement of $S_L$.
Keeping both $S_R$ and $S_L$, Belle had to assume
reality to make a 2D plot~\cite{Belle-B2munu_Morio19}.
But, sourced in Yukawa couplings, they are clearly complex.
%

%
The correspondence with extra Yukawa couplings is,
\begin{eqnarray}
 S_L^{\ell'\ell} = \frac{m_b(\mu_b)}{m_b(\mu_0)}
                            \frac{v^2}{2m_{H^+}^2}
                            \rho_{\ell'\ell}^*\rho_{tu}^*
                            \frac{V_{tb}}{V_{ub}},
\label{eq:SLlpl}
\end{eqnarray}
where $\mu_0$ is at $m_{{H^+}}$ scale.
We note that Yukawa couplings are dimension-4 terms
in the Lagrangian. 
For leptonic $B^+$ decay, QCD correction is easy to match
with 4-Fermi operators. But it was the 
insight on $|\rho_{bb}|$ vs $|\rho_{tu}^*V_{tb}/V_{ub}|$
that allowed us to drop the $S_R$ term.

Ignoring $\ell' = e$, i.e. taking $\rho_{e\ell}$ as negligible,
we get, 
\begin{eqnarray}
 && {\cal B}(B \to \mu \bar\nu) = {\cal B}(B \to \mu \bar\nu_\mu)|^{\rm SM} \nonumber\\
 && \; \times \left(
        \left| 1 - \frac{m_{B}^2}{m_b m_\mu} S_L^{\mu\mu}\right|^2
     + \left|      \frac{m_{B}^2}{m_b m_\mu} S_L^{\tau\mu}\right|^2
   \right),
\label{eq:Bqmunu}
\end{eqnarray}
where the $S_L^{\mu\mu}$ effect from diagonal 
$\rho_{\mu\mu}$ coupling interferes with SM (the ``1''),
while the $S_L^{\tau\mu}$ effect from off-diagonal 
$\rho_{\tau\mu}$ adds in quadrature.
The {$\tau$ equivalent of the} former can be found in 
an early $B\to \tau\nu$ study~\cite{Crivellin:2012ye},
but the more detailed follow-up~\cite{Crivellin:2013wna}
 erroneously summed over $\ell'$ in amplitude,
incorrectly giving neutrino flavor independence{,
hence missing the second effect.
Furthermore, Ref.~\cite{Crivellin:2013wna}
kept $\tan\beta$ in the formulation,
but $\tan\beta$ is not physical in g2HDM~\cite{untbeta}.
In fact, $B \to \mu\bar\nu$ has not been emphasized in the literature.
The one paper that did, Ref.~\cite{Chen:2018hqy},
not only followed Ref.~\cite{Crivellin:2013wna} in keeping $\tan\beta$,
but {\it assumed} the charged lepton Yukawa matrix 
$\rho^\ell$ to be diagonal, thereby leaving out our second term.}

The $S_L^{\tau\mu}$ effect has {thus} not been discussed before.
Without interference it may appear less important,
but we will show it is the leading effect.
%
%
For $B \to \tau\bar\nu$,
one simply exchanges $\mu \leftrightarrow \tau$ 
in Eq.~(\ref{eq:Bqmunu}).
%
%

We parametrize
$S_L^{\ell'\ell} = \bigl| S_L^{\ell'\ell} \bigr|\,e^{i\phi_{\ell'\ell}}$,
where $\phi_{\ell'\ell}$ is the phase difference between 
$\rho_{\ell'\ell}^*\rho_{tu}^*$ and $V_{ub}$.
The phase does not enter the off-diagonal effect,
but the two mechanisms must be treated separately.
Setting $S_L^{\tau\mu} = 0$ and taking $\overline{m_b}(\mu_b) = 4.18$ GeV 
in Eq.~(\ref{eq:Bqmunu}), 
Fig.~\ref{fig:fig1}[left] illustrates Eq.~(\ref{eq:B2munu-Belle19})
in the $\bigl | S_L^{\mu\mu} \bigr|$--$\phi_{\mu\mu}$ plane, where
 dotted (red solid) line is the Belle central value
 (SM expectation of $3.92 \times 10^{-7}$), with two different shades
illustrating $\pm 1\sigma$ and $\pm 2\sigma$ ranges. 
The plot is symmetric for $\phi_{\mu\mu} < 0$.
%

The in-quadrature second term of Eq.~(\ref{eq:Bqmunu}) 
is $\phi_{\tau\mu}$-independent. Setting  $S_L^{\mu\mu} = 0$,
we plot in Fig.~\ref{fig:fig1}[right] ${\cal B}(B \to \mu\bar\nu)$
 vs $\bigl | S_L^{\tau\mu} \bigr|$, which is the blue dashed line. 
As there is only enhancement,
the gray area below SM expectation 
is inaccessible.
Belle data constrain $\bigl | S_L^{\tau\mu} \bigr| \lesssim 0.019$,
the same as imaginary $S_L^{\mu\mu}$ in Fig.~\ref{fig:fig1}[left].
Setting $S_L^{\mu\tau} = 0$, Fig.~\ref{fig:fig3}[left] depicts in 
the $\bigl | S_L^{\tau\tau} \bigr|$--$\phi_{\tau\tau}$ plane 
the Belle average ${\cal B}(B \to \tau\bar\nu)
 \simeq (9.1 \pm 2.2) \times 10^{-5}$ from PDG,
with notation analogous to Fig.~\ref{fig:fig1}.
Likewise, Fig.~\ref{fig:fig3}[right] plots ${\cal B}(B \to \tau\bar\nu)$
vs  $|S_L^{\mu\tau}|$ with $S_L^{\tau\tau} = 0$.
The bands in Fig.~\ref{fig:fig3} appear narrower
not so much as an artifact of plotting, but 
reflects ${\cal B}(B \to \tau\bar\nu)$ being 
better measured than ${\cal B}(B \to \mu\bar\nu)$.

\paragraph{Interpretation in g2HDM.---}
%

Interpreting in g2HDM that plausibly holds fundamental 
extra Yukawa couplings, sheds light on the underlying physics.

${\cal B}(B \to \mu\bar\nu)$ from Belle, 
Eq.~(\ref{eq:B2munu-Belle19}), is consistent with SM, 
which Belle~II would measure in due time~\cite{Kou:2018nap}.
But if one has $+1\sigma$ or even $+2\sigma$ enhancement 
over the central value, then earlier discovery is possible.
From Fig.~\ref{fig:fig1} we find 
$+1\sigma$ to $+2\sigma$ enhancement correspond to
$|S_L^{\mu\mu}| \in (0.006, 0.009)$ for the constructive (negative) case,
 (0.015, 0.019) for the imaginary case, and
 (0.038, 0.041) for the destructive (positive) case.
The incoherent, second effect of Eq.~(\ref{eq:Bqmunu})
has the same parameter range as imaginary $S_L^{\mu\mu}$,
but for $|S_L^{\tau\mu}| \in (0.015, 0.019)$, 
i.e. with $\bar\nu_\tau$ index.

We take $m_{{H^+}} = 300$ GeV as benchmark
 for sake of the largest effect, 
but also because the usual $m_{{H^+}}$ bound
 from 2HDM~II does not apply
 (some discussion can be found in Refs.~\cite{Altunkaynak:2015twa}
  and \cite{Hou:2019grj}), 
given the many new flavor parameters.
From Eq.~(\ref{eq:SLlpl}) we find
%
$|S_L^{\ell'\ell}| \simeq 150\bigl |\rho_{\ell'\ell}  \rho_{tu}\bigr| 
  (300\, {\rm GeV}/m_{{H^+}} )^2$.
%
The 
constructive case of  
$S_L^{\mu\mu} \in (-0.009, -0.006)$ needs 
$ |\rho_{\mu\mu}\rho_{tu}| \simeq (4$--6)$ \times 10^{-5}$,
and would grow as $(m_{{H^+}}/{\rm 300\; GeV})^2$.
So, what do we know, or can infer, about
 $ |\rho_{\mu\mu}|$ and $ |\rho_{tu}|$?
Given that $H^+$ effect is normalized to SM,
Eq.~(\ref{eq:Mod})  offers a clue:
$\rho_{\mu\mu}^*$ is ``normalized'' against $\lambda_\mu \simeq 0.0006$,
the charged lepton Yukawa coupling,
while $\rho_{tu}^*$ is normalized to $\hat m_b \sim 0.015$, 
the ``effective Yukawa coupling'' from $m_b$ evaluated at $m_{{H^+}}$ scale.
The combined $\lambda_\mu \hat m_b \sim 1 \times 10^{-5}$
suggests $S_L^{\mu\mu}$ 
falls short of
enhancing ${\cal B}(B\to \mu\bar\nu)$ even for the most optimistic case, 
let alone the larger $|S_L^{\mu\mu}|$ 
needed for imaginary or destructive cases.

It is instructive to take a look at the second mechanism, 
i.e. via $\bar\nu_\tau$ flavor.
$|S_L^{\tau\mu}| \in (0.015, 0.019)$ is the same as 
imaginary $S_L^{\mu\mu}$,
hence $|\rho_{\tau\mu}\rho_{tu}| \gtrsim 10^{-4}$ 
may not appear promising. 
However, 
due to the hint for $h \to \tau\mu$ in CMS Run~1 data,
up until early 2017, values of $|\rho_{\tau\mu}|$ 
as large as 0.26 had been entertained.
Although the hint disappeared with Run~2 data, 
it could reflect approximate alignment, 
i.e. a small $\cos\gamma$. 
As discussed below, if we allow $\rho_{\tau\mu} \sim \lambda_\tau$,
then 
$\lambda_\tau \hat m_b \sim 1.5\times 10^{-4}$
seems to allow the $\rho_{\tau\mu}$ mechanism to 
enhance $B\to \mu\bar\nu$. 

Taking a closer look, 
%
we {suggest} $\rho_{\mu\mu} = {\cal O}(\lambda_\mu)$ 
{is} reasonable, 
as $\lambda_\mu$ arises from diagonalizing the mass matrix,
but $\rho_{\mu\mu}$ is from an orthogonal combination of
the two unknown Yukawa matrices, 
going through the same diagonalization. 
To avoid fine tuning, these two Yukawa matrices must each contain
the ``flavor organization''~\cite{Hou:2017hiw} reflected in mass-mixing hierarchies,
hence $\rho_{\mu\mu} = {\cal O}(\lambda_\mu)$.

We treat  $\rho_{\tau\mu}$ more liberally, as argued above 
for 
$|\rho_{\tau\mu}| \sim \lambda_\tau \sim 0.01$,
since much larger  $\rho_{\tau\mu}$ values have been considered only recently.
The most relevant constraint comes from $\tau \to \mu\gamma$,
where the two-loop mechanism constrains 
$|\rho_{\tau\mu}| \lesssim 0.01$~\cite{Hou:2019grj}
for $\rho_{tt} \sim 1$, but 
weakens for weaker $\rho_{tt}$.
Finally, having $|\rho_{\tau\mu}| \lesssim |\rho_{\tau\tau}| \sim \lambda_\tau$
 is not unreasonable, just as 
$|\rho_{tc}|$ could be up to $|\rho_{tt}| \sim \lambda_t$~\cite{Fuyuto:2017ewj},
where $\rho_{tt}$ and $\rho_{tc}$ provide 
two possible CP violating sources for electroweak baryogenesis, 
which strongly motivates g2HDM.
Thus, we suggest $|\rho_{\tau\mu}| \lesssim 0.02$ as reasonable,
and its value is in any case an experimental issue.

For the common $\rho_{tu}$ factor, things are harder to discern.
Taking $|\rho_{tu}| \sim \sqrt{2m_tm_u}/v \sim 0.003$
would be a bit small, but it need not be that small, 
since the direct $t \to uh$ search bound~\cite{Aaboud:2018oqm}
is not so different from $t \to ch$, hence 
quite forgiving.
In lack of a true yardstick, we take $|\rho_{tu}| \lesssim \hat m_b$ as reasonable. 

Thus, even taking
 $|\rho_{\mu\mu}| \sim 3\lambda_\mu$ 
 and $|\rho_{tu}| \sim 2\hat m_b$, 
$|\rho_{\mu\mu}\rho_{tu}| \sim 5 \times 10^{-5}$
is only borderline in enhancing ${\cal B}(B\to \mu\bar\nu)$ 
for the most optimistic, constructive case, 
and in general would not quite suffice. 
However, even modest 
$|\rho_{\tau\mu}| \lesssim \lambda_\tau$ and $|\rho_{tu}| \lesssim \hat m_b$
give $|\rho_{\tau\mu}\rho_{tu}| \lesssim 1.5 \times 10^{-4}$,
allowing reasonable outlook for enhancement 
even 
if it comes only in quadrature.
For higher $m_{{H^+}}$, e.g. 500--600 GeV, $|\rho_{\tau\mu}|$ and $|\rho_{tu}|$ 
in the upper reaches of our suggested range can 
still 
give enhancement .

%
%

Turning to $B \to \tau\bar\nu$, we take
$\rho_{\tau\tau} = {\cal O}(\lambda_\tau)$
and $\rho_{\mu\tau} \lesssim \lambda_\tau$.
For constructive case, 
we see from Fig.~\ref{fig:fig3}[left] that 
$S_L^{\tau\tau} \in (-0.065,\, -0.035)$
 for $+1\sigma$ to $+2\sigma$ enhancement, 
which suggests 
$|\rho_{\tau\tau} \rho_{tu}| \simeq (2.3$--$4.3) \times 10^{-4}$
for $m_{{H^+}} \simeq 300$ GeV.
But $|\rho_{\tau\tau} \rho_{tu}|
 \sim \lambda_\tau \hat m_b \sim 1.5 \times 10^{-4}$
 falls short. 
Enhancement from SM is possible in constructive case 
only for $|\rho_{\tau\tau} \rho_{tu}|$ in the 
upper reaches of $\sim 6\lambda_\tau \hat m_b$,
but gets easily damped by larger $m_{{H^+}}$.
%
For the second effect, Fig.~\ref{fig:fig3}[right]
suggests $|S_L^{\mu\tau}| \in (0.15,\, 0.2)$, 
much larger than the constructive case.
With $\rho_{\mu\tau} \rho_{tu} \lesssim \lambda_\tau\hat m_b$, 
this mechanism cannot enhance $B \to \tau\bar\nu$.

Thus, one {\it expects} ${\cal B}(B \to \tau\bar\nu)$ in g2HDM 
to be 
SM-like,
%
while ${\cal B}(B \to \mu\bar\nu)$
{\it could} be better enhanced. 

\paragraph{Discussion.---}
%
%
$K \to \mu\bar\nu$ decay is not constraining, 
as both coherent and incoherent effects are suppressed by
$|V_{ts}V_{ub}/V_{tb}V_{us}| \, (m_K^2\hat m_b/m_B^2\hat m_s) \sim 0.0003$, 
while $K \to e\bar\nu$ is even more SM-like.
The same argument goes with pion decays,
%
and the effect in $D^+$, $D_s$ decays is also rather weak.
%
For $B_c$, 
we do not see how $B_c \to \ell\bar\nu$ 
can be reconstructed.
Thus, $B \to \mu\bar\nu$ provides the unique probe
of extra Yukawa couplings in g2HDM,
whereas $B\to \tau\bar\nu$ is expected to be SM-like.
%
%
Taking the ${\cal R}_B^{\mu/\tau}$ ratio eliminates the main
uncertainties associated with $|V_{ub}|$. 
It is interesting that the $S_L^{\mu\mu}$ mechanism could also 
{\it suppress} ${\cal B}(B \to \mu\bar\nu)$
(see lower left region of Fig.~\ref{fig:fig1}[left]),
but would take longer for Belle~II to uncover.
{We note in passing that a deviation in ${\cal B}(B \to \mu\bar\nu)$
may also be caused by leptoquarks~\cite{LQ}
($W'$ is overly constrained).}

What about $\mu \to e\nu\bar\nu$ and $\tau \to \ell\nu\bar\nu$ decays?
As these are dominated by $V-A$ theory,
the vector currents couple via $g \sim {\cal O}(1)$,
without helicity suppression.
In contrast, since 
$|\rho_{\tau\mu}| \lesssim |\rho_{\tau\tau}|
 = {\cal O}(\lambda_\tau) \ll g$
are the largest Yukawa couplings that enter,
 together with $M_W^2/m_{{H^+}}^2$ suppression, 
Nature has quite an effective mechanism in
hiding the extra Yukawa coupling effects in the lepton sector.
For example, given the extreme lightness and abundance of the electron,
$\rho_{e\mu}$ and $\rho_{e\tau}$ must be very small,
we expect $\mu \to e\nu\bar\nu$ and $\tau\to e\nu\bar\nu$
to be SM-like to high precision.
Similar arguments hold for
$B \to X_u \ell\bar\nu$, $\pi\ell\bar\nu$ decays,
which are plagued further by hadronic uncertainties.
%
{Finally, we have used $\tau \to \mu\gamma$ to constrain $\rho_{\tau\mu}$.
As a crosscheck, 
for $\rho_{\tau\mu} \lesssim 2\lambda_\tau \simeq 0.02$,
   $\rho_{\mu\mu} \lesssim 3\lambda_\mu \simeq 0.0018$,
 and $m_{A^0} \gtrsim 300$ GeV
 (degenerate with $H^+$, ignoring heavier $H^0$,
 and approximate alignment control of $h^0$ effect),
we estimate ${\cal B}(\tau \to \mu\mu\mu) \lesssim {\cal O}(10^{-11})$,
which is far below current~\cite{Tanabashi:2018oca} experimental bound.}

Nature does hide well the effect of extra Yukawa couplings 
in $H^+$ mediated low energy processes. 
$B \to \mu\bar\nu$ is more helicity suppressed than $B \to\tau\bar\nu$,
with $\rho_{\tau\mu}$ giving $\mu\bar\nu_\tau$ final state,
and $b \to u$ transition 
giving $V_{tb}/V_{ub}$ enhancement of $\rho_{tu}$,
both of which can happen only in g2HDM.
Our imprecise knowledge of 
{$\rho_{\tau\mu}$ and $\rho_{tu}$}
 allow for enhancement: 
 ${\cal B}(B \to \mu\bar\nu)$ {\it probes the 
extra Yukawa coupling 
{product}
 $\rho_{\tau\mu}\rho_{tu}$}.
But early 
impressions of enhanced $B \to \tau\nu$~\cite{Tanabashi:2018oca} 
trained people to expect NP in $B \to \tau\nu$, 
as reflected in the Belle~II Physics Book~\cite{Kou:2018nap}.
%

\paragraph{Conclusion.---}
%

With a second Higgs doublet quite plausible, 
the existence of extra Yukawa couplings is an experimental issue.
The SM and 2HDM~II predict the ratio ${\cal R}_B^{\mu/\tau} =
 {\cal B}(B \to \mu\bar\nu)/{\cal B}(B \to \tau\bar\nu)$ 
to be 0.0045, which offers a unique test.
Through $\bar\nu_\tau$ flavor, 
the $\rho_{\tau\mu}$ coupling can enhance $B \to \mu\bar\nu$, 
while $B \to \tau\bar\nu$ is SM-like.
%
%
If enhancement of ${\cal R}_B^{\mu/\tau}$
is uncovered by Belle~II with just a few ab$^{-1}$, then 
the many extra Yukawa couplings --- fundamental flavor parameters 
associated with a second Higgs doublet --- would need to be unraveled.

\vskip0.2cm
\noindent{\bf Acknowledgments.} \
We thank A. Crivellin for communications.
This research is supported by grants MOST 106-2112-M-002-015-MY3,
107-2811-M-002-3069, 107-2811-M-002-039, 
and NTU 108L104019.


\end{document}